\documentclass[12pt,preprint]{aastex}

\newcommand{\kms}{km~s$^{-1}$}
\newcommand{\cm}{cm$^{-2}$}
\newcommand{\lya}{{\rm Ly}$\alpha$}

\slugcomment{Submitted to ApJ}
\shorttitle{Heavy Elements and Dust in three GRBs}
\shortauthors{Savaglio, Fall, Fiore}

\begin{document} 

\input epsf 

\title{Heavy Element Abundances and Dust Depletions \\
in the Host Galaxies of Three  Gamma--Ray Bursts}

\author{Sandra Savaglio\altaffilmark{1}}
\affil{Johns Hopkins University, 3400 North Charles Street,
Baltimore, MD21218; savaglio@pha.jhu.edu}

\author{S. Michael Fall}
\affil{Space Telescope Science Institute, 3700 San Martin Drive, 
        Baltimore, MD21218; fall@stsci.edu}

\author{Fabrizio Fiore}
\affil{Osservatorio Astronomico di Roma, via di Frascati 33, 
Monteporzio, I00040, Italy; fiore@quasar.mporzio.astro.it}

\altaffiltext{1}{On leave of absence from Osservatorio Astronomico di
Roma, Italy} 

\begin{abstract} 

We have derived the column densities of heavy elements in three
gamma--ray burst (GRB) optical transients, associated with the
circumburst or interstellar medium (ISM) of the host galaxy. In
comparison with the same elements observed in damped Lyman--$\alpha$
(DLA) systems along QSO sight lines, we find evidence for much
higher column densities of \ion{Zn}{2}.  The gap between the
QSO--DLA and GRB--DLA distributions is smoothly bridged by
observations of the interstellar absorption in the Milky Way and the
Magellanic Clouds.  Very small [Fe/Zn], [Si/Zn], and [Cr/Zn] values
in GRB--DLAs indicate large dust depletions. Once the dust--to--metals
ratios are determined, we find an optical extinction $A_V\approx1$
mags, to be compared with typical $A_V\lesssim0.1$ in most QSO--DLAs.
Our inference of high dust content appears to be in contradiction
with the typical low reddening previously found in GRBs. One possible
way to reconcile is a dust grain--size distribution biased towards big
grains, which would give a grey extinction. Possibly the small dust
grains have been destroyed by the GRBs themselves. Our findings
support the idea that primarily optically selected QSOs probe mainly
low gas/dust regions of high redshift galaxies, while the more
powerful GRBs can be detected through denser regions of their ISM
(molecular clouds and star forming regions). Therefore GRB--DLAs and
QSO--DLAs together provide a more complete picture of the global
properties of the interstellar medium in high redshift galaxies.

\end{abstract}


\section{INTRODUCTION}

Gamma--ray bursts (GRBs) are among the most dramatic events in the
Universe. Their luminosities are so high ($10^{51} - 10^{54}$ ergs are
emitted in a few seconds) that they can be seen at the highest
redshifts. However, since the decline in their energy emission is very
fast, the investigation and understanding of their physical nature is
is not particularly easy (Piran 2001).  For instance, the redshift of
the afterglow or host galaxy has been measured in only 24
cases\footnote{see http://www.mpe.mpg.de/$\sim$jcg/grbgen.html for the
complete list of GRB redshifts}.

Only eight of these objects have been targeted quickly enough by
ground--based telescopes to obtain low resolution spectra of the
optical transient, and all show a common peculiarity: very strong UV
absorption lines of low ionization species (Fiore 2001; Fynbo et
al.~2002). This is similar to what is seen in damped Lyman--$\alpha$
systems (DLAs) associated with intervening galaxies along QSO sight
lines. GRB--DLAs are certainly an unexplored and important source of
information on the circumburst and/or interstellar medium (ISM) of the
GRB host galaxy.  Here we explore in more detail their properties and
compare them with those of QSO--DLA galaxies.

We use the curve of growth (COG) analysis to derive column
densities of low ionization species in several GRB--DLAs. As for
QSO--DLAs, we neglect ionization effects to derive the abundances of
several heavy elements. The low ionization of the gas is warranted by
the large equivalent widths of the low ionization lines, and the low
equivalent widths of the moderate and high ionization lines. The
difference in abundances between weakly and greatly dust depleted
elements provides an estimate of the dust content.

The COG analysis was applied to QSO--DLAs (Blades et al.~1982;
Turnshek et al.~1989) before the advent of high resolution
spectroscopy, which allows a direct fitting of individual absorption
lines. Since GRB emission fades very rapidly, high resolution (FWHM
$\sim 10$ \kms) spectra of GRB afterglows are very hard to obtain.

In this work, we interpret strong differences between GRB--DLAs and
QSO--DLAs as the result of the difference in the densities of the
intervening gas. Since GRBs are, at least for some time, brighter than
primarily optically selected QSOs, they can be observed even if their
sight lines cross very dense and dusty regions, such as molecular
clouds, where extinction is very severe. On the other hand, QSO sight
lines are easier to detect when intersecting less dusty interstellar
clouds.

This bias in QSO--DLAs has been already suggested in the past (see,
for instance, Fall \& Pei 1993; Boiss\'e et al.~1998). More
recently, Ellison et al.~(2001) have studied a small sample of
radio--selected QSOs and tentatively found more DLAs than in optically
selected QSOs. Even if this result is affected by small number
statistics (significant at $\sim 2 \sigma$ level), it supports the
idea of dust obscuration in optically selected QSO--DLAs.

Our analysis is alternative to and independent of the dust
content estimate obtained so far for GRBs.  The dust reddening in GRB
afterglows has been derived in several objects (see for instance
Reichart 1998; Jensen et al.~2001; Rhoads \& Fruchter 2001; Fynbo et
al.~2001; Galama \& Wijers 2001) and in most cases it has been found
to be small or negligible.  The method applied assumes initial power
law emission from the GRBs, and steep extinction curves such as those
in the Milky Way (MW) or Large \& Small Magellanic Clouds
(MC). Possible discrepancies with the results of our analysis
challenge the two initial assumptions of very steep extinction and
initial power law emission of the central source.

For this study, we examine the absorption lines of three of the
eight GRBs with UV rest frame spectra, namely GRB990123 ($z=1.601$,
Kulkarni et al.~1999), GRB010222 ($z=1.475$, Masetti et al.~2001; Jha
et al.~2001; Salamanca et al.~2001; Mirabal et al.~2002), and
GRB000926 ($z=2.0379$, Castro et al.~2001). We exclude GRB970508
($z=0.835$, Metzger et al.~1997), GRB990510 ($z=1.619$, Vreeswijk et
al.~2001) GRB000301C ($z=2.04$, Jensen et al.~2001), GRB990712
($z=0.4331$, Vreeswijk et al.~2001), and GRB011211 (Holland et
al.~2002) for poor quality of the spectra and therefore lack of
interesting absorption features.

\section{GRB--DLAs COLUMN DENSITY DETERMINATION}

Similarly to QSO--DLAs, GRB--DLAs show strong \ion{Fe}{2},
\ion{Mg}{2}, and \ion{Si}{2} absorptions, together with relatively
weak \ion{Zn}{2} and \ion{Cr}{2}. Table \ref{t1} lists oscillator
strengths $f_\lambda$ (from the recent compilation by Prochaska et
al.~2001) and rest frame equivalent widths $W_r$ of absorption lines
in the three GRBs used in the present analysis.  The large $W_r$ of
all these species relative to higher ionization species like
\ion{C}{4}, \ion{Si}{3} and \ion{Si}{4}, indicate that the medium is
weakly ionized. From the equivalent widths (EWs), we derive column
densities and directly compare the results with typical column
densities in QSO--DLAs. Comparing column densities instead of
metallicities [X/H] has the advantage that the \ion{H}{1} column
density (known only in GRB000926 with large uncertainty; Fynbo et
al.~2002) is not required. Moreover, it is not necessary to make any
assumptions about the dust depletion, cosmic abundance or ionization
level of the considered element, but we can still search for
similarities and differences between GRB--DLAs and QSO--DLAs.

A direct comparison is complicated by the fact that normally QSO--DLAs
are observed at high spectral resolution (FWHM $\simeq10$ \kms) in a
large wavelength range (thanks to echelle spectroscopy), allowing a
good determination of column densities of many ions, even in cases
where saturated or weak lines in complex features are detected (see
for instance Lu et al.~1996; Pettini et al.~1997; Prochaska \& Wolfe
1999). On the other hand, the typical resolution of GRB optical
spectra is FWHM $>100$ \kms. Moreover, the majority of absorption
lines are saturated, and it is not possible to use the linear part of
the COG to determine column densities.  However, the lines are
numerous and so, even if the uncertainty may be large, it is possible
to make an abundance analysis using the general COG analysis (Spitzer
1978). Since the typical resolution of GRB spectra does not allow the
separation of the different components of a complex feature, due to
the many clouds along the line of sight, in our analysis we adopt a
large ``effective'' Doppler parameter, being the result of the
superposition of many narrow absorption lines. For reference, we note
that typical velocities in QSO--DLAs are $\sim 100$ \kms~or more. The
COG analysis applied with the single component approximation was used
for QSO--DLA studies before high resolution spectroscopy of QSOs
became possible (see for instance Blades et al.~1982; Turnshek et
al.~1989). In fact, Jenkins (1986) showed that the COG technique
applied to complex features gives nearly the correct answer (the
simulated--to--true column density ratio rarely goes below 0.8) even
if different lines have very different saturation levels or Doppler
parameters.  This general result is also confirmed by Savage \&
Sembach (1991).

\begin{figure}
\centerline{{\epsfxsize=17.cm \epsfysize=17.cm \epsfbox{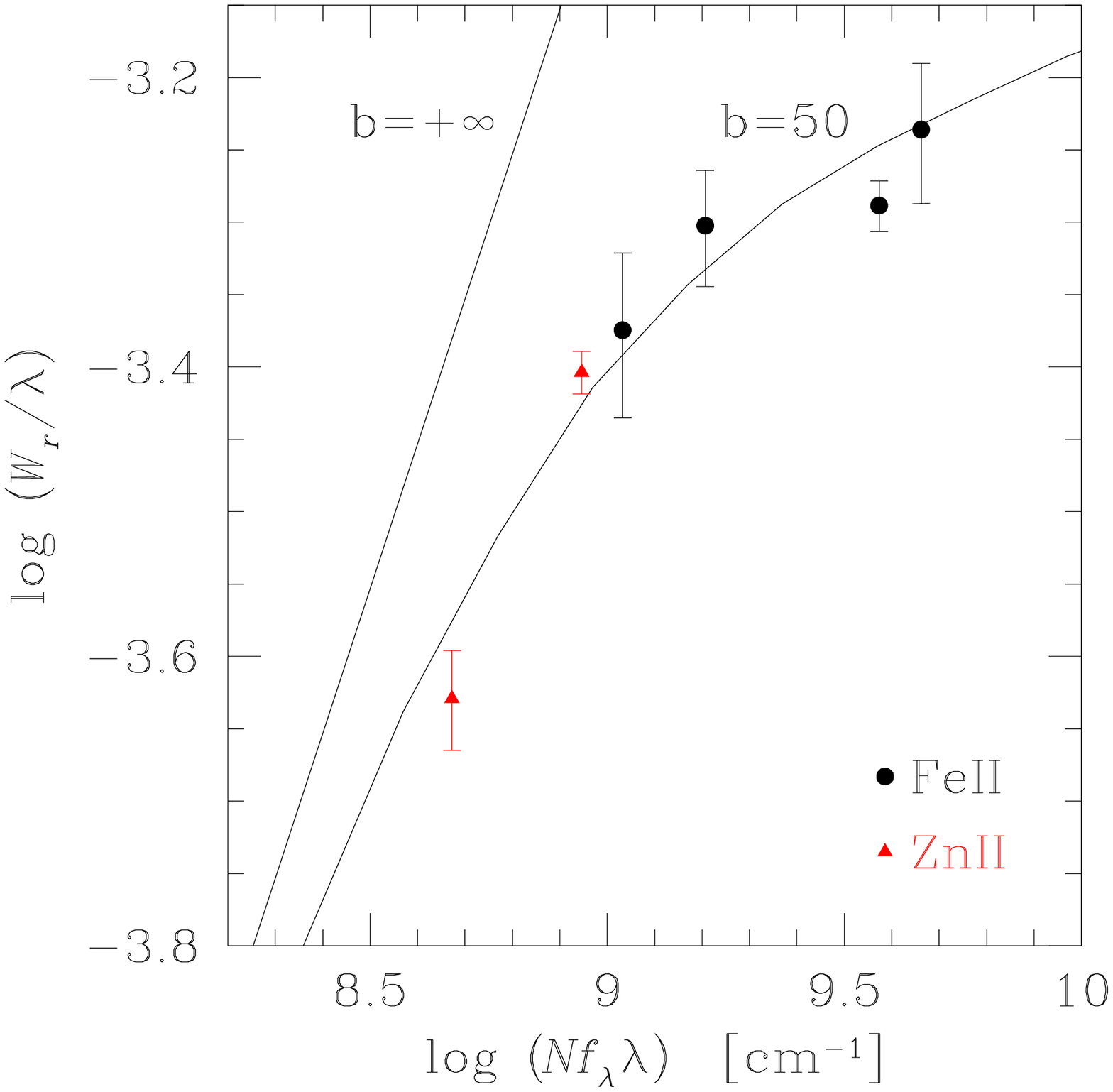}}}
\caption[f1]{Curve of growth for GRB990123 absorption lines ($b$ is
the effective Doppler parameter in \kms). Points give the best--fit
column densities, reported in Table \ref{t2}. The straight line is the
linear case ($b=+\infty$).}
\label{f1a}
\end{figure}

\begin{figure}
\centerline{{\epsfxsize=17.cm \epsfysize=17.cm \epsfbox{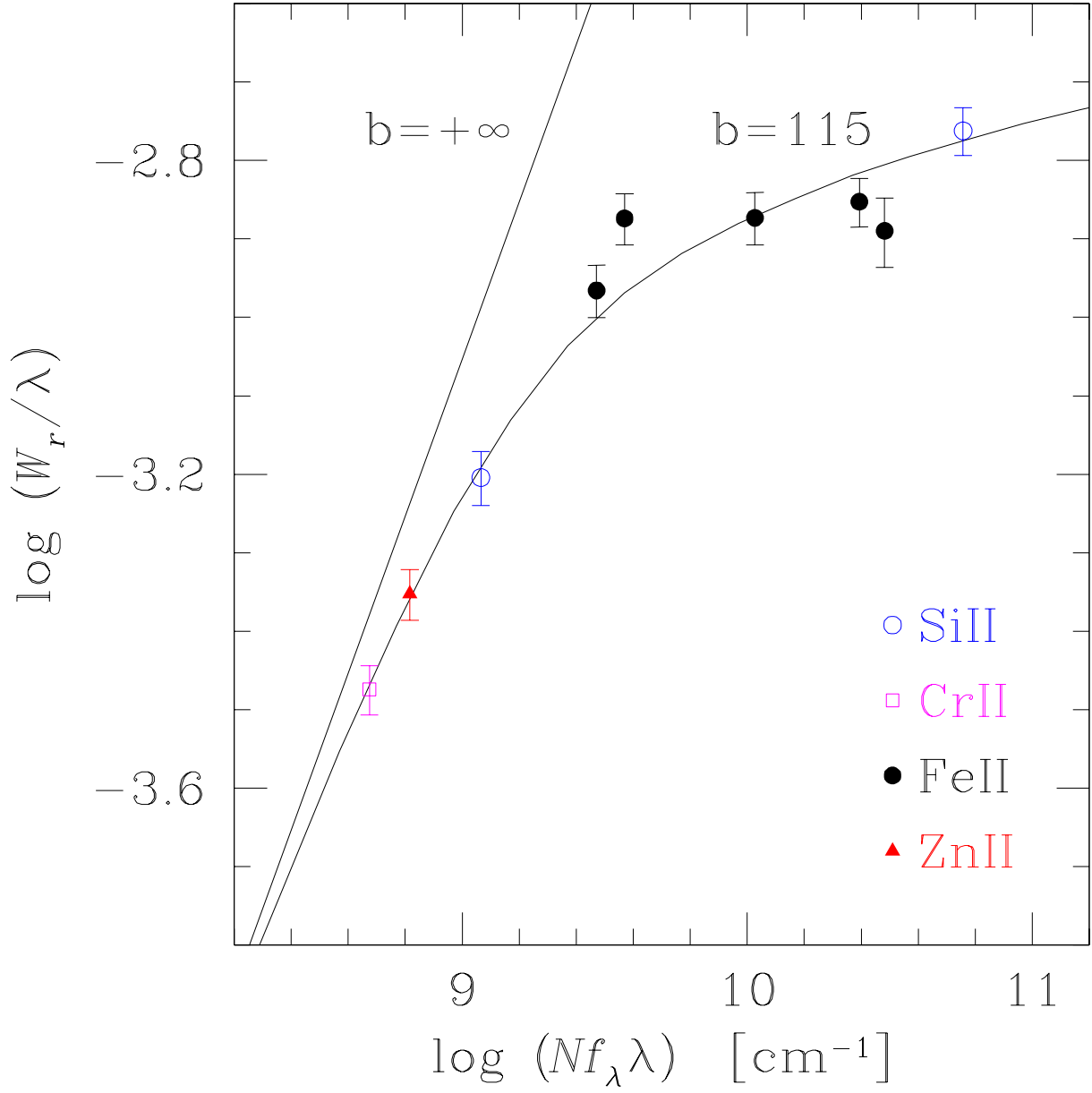}}}
\caption[f1]{Same as in Figure \ref{f1a} but for GRB000926 absorption lines.}
\label{f1b}
\end{figure}

\begin{figure}
\centerline{{\epsfxsize=17.cm \epsfysize=17.cm \epsfbox{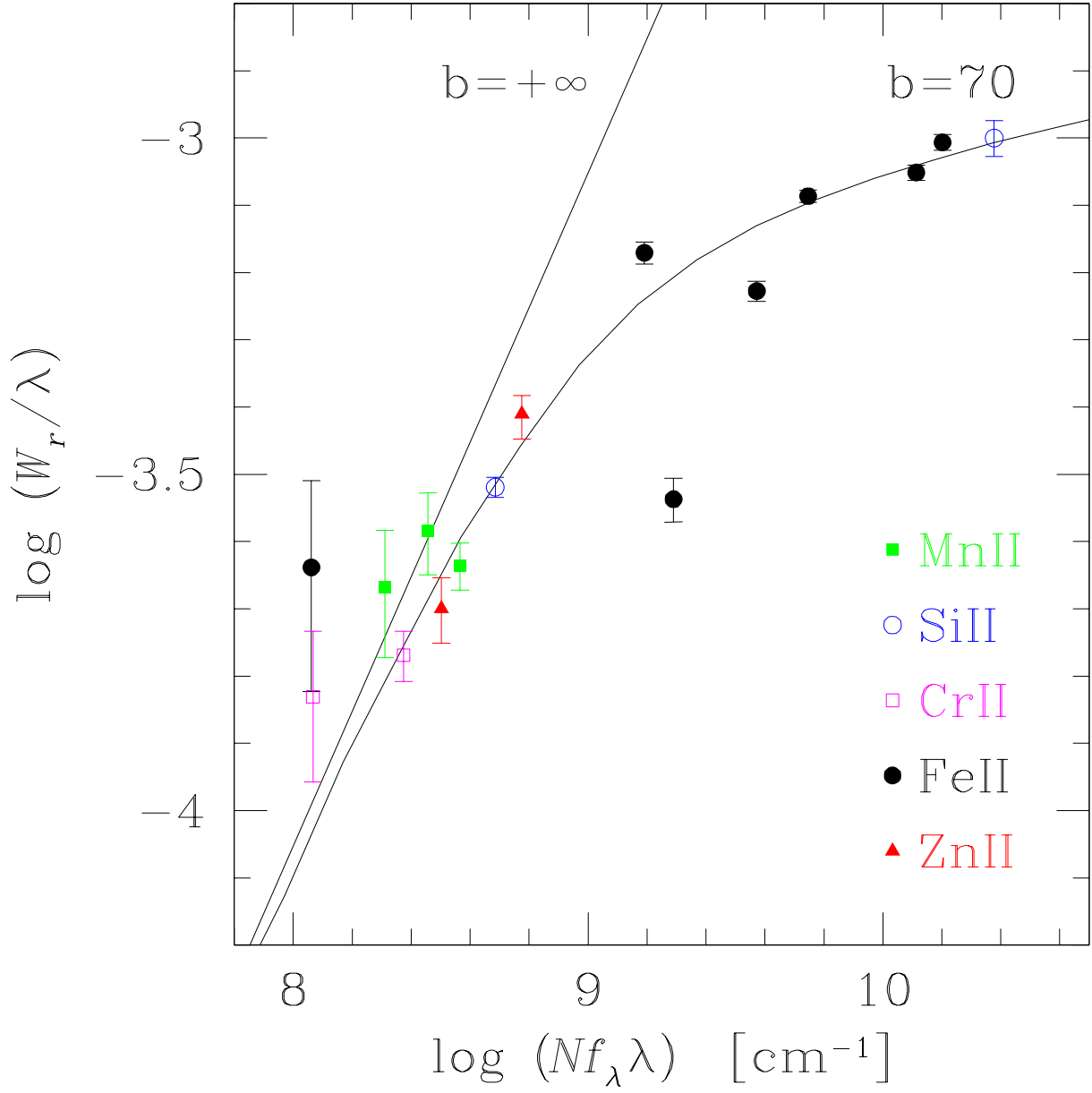}}}
\caption[f1]{Same as in Figure \ref{f1a} but for GRB010222 absorption
lines.}
\label{f1c}
\end{figure}

The best--fitting column density is determined by using simultaneously
different EWs of various ions and choosing the COG with an effective
Doppler parameter that minimizes the $\chi^2$. Once the best effective
Doppler parameter is determined, the error on the column density for
each ion is determined by varying $\log N$ for which $\chi^2/dof
\lesssim 1$.  In the following \S~2.1--2.3, we discuss column density
measurements in each individual GRB. Figures \ref{f1a}--\ref{f1c} show
the best--fitting $W_r/\lambda$ and column densities, and the
corresponding COG.  Estimated column densities are given in Table
\ref{t2}.

\subsection{GRB990123}

The optical transient of GRB990123 at $z=1.6004$ was observed with the
Keck/LRIS (Kulkarni et al.~1999) for a final resolution and
signal--to--noise ratio of FWHM $=11.6$ \AA~($400-700$ \kms) and S/N
$\approx30$, respectively.  The very weak \ion{Fe}{2} $\lambda2260$ EW
gives inconsistent results with other \ion{Fe}{2} lines, and thus this
detection has not been considered in our analysis.  \ion{Fe}{2}
$\lambda2374$ is not reliable because it is probably partly blended
with the \ion{Fe}{2} $\lambda$2382 line.

For the four remaining \ion{Fe}{2} lines, we get a best--fit of $\log
N($\ion{Fe}{2}$)=14.78^{+0.17}_{-0.10}$ and $b=50$ \kms (Figure
\ref{f1a}).  From the EW of \ion{Zn}{2} $\lambda2026$ and considering
$b=50$ \kms, we derive a \ion{Zn}{2} column density of $\log
N($\ion{Zn}{2}$) \simeq 14$, and calculate a \ion{Zn}{2} $\lambda2062$
contribution to the \ion{Zn}{2} $\lambda2062$ + \ion{Cr}{2}
$\lambda2062$ blend of $W_r = 0.58$ \AA, to be compared with the
detected $W_r = 0.48\pm0.04$ \AA.  This means that the contribution of
the \ion{Cr}{2} $\lambda2062$ line to the blend is negligible or non
existing. Therefore we recalculate the \ion{Zn}{2} column density from
the \ion{Zn}{2} $\lambda2026$ and \ion{Zn}{2} $\lambda2062$ doublet,
from which we derive $\log N($\ion{Zn}{2}$)=13.95^{+0.05}_{-0.05}$

\subsection{GRB000926}

GRB000926 at $z=2.0379$ was observed with the Keck/ESI echelle
spectrograph (Castro et al.~2001), with a good signal--to--noise ratio
(S/N $=10-20$) and relatively high resolution (FWHM $\approx 80$
\kms). Castro et al.~(2001) also show a low resolution spectrum, with
EW measurements that are systematically larger than those obtained
from the high resolution spectrum, but the authors claim that the high
resolution measurements are more accurate, so we only use these.  The
spectrum shows at least two absorbing clouds separated by 168 \kms. In
our analysis, we consider the total equivalent widths of the two
components together, since the resolution does not allow them to be
clearly separated.

The observed \ion{Si}{2} $\lambda1808$ and \ion{Si}{2} $\lambda1526$
lines, and the five detected \ion{Fe}{2} lines give $\log
N($\ion{Si}{2}$)=16.47^{+0.10}_{-0.15}$ and $\log
N($\ion{Fe}{2}$)=15.6^{+0.20}_{-0.15}$, respectively ($b=115$ \kms,
Figure \ref{f1b}).  If we assume $b=115$ \kms, we get $\log
N($\ion{Cr}{2}$) = 14.34^{+0.05}_{-0.05}$ from the \ion{Cr}{2}
$\lambda2056$ EW.  This corresponds to $W_r=0.5-0.6$ \AA~and
$W_r=0.35-0.43$ \AA~ for \ion{Cr}{2} $\lambda2062$ and \ion{Cr}{2}
$\lambda2066$, respectively.  Neither of these measurements is
consistent with those observed ($W_r<0.28$ \AA~and $W_r<0.2$ \AA,
respectively); this might indicate a contamination of the \ion{Cr}{2}
$\lambda2056$ line by an unidentified absorption associated with an
intervening metal system.  Moreover, the \ion{Cr}{2} $\lambda2062$ is
probably corrupted by a noise spike in the spectrum (see Figure 3
of Castro et al.~2001) probably due to a remnant of the strong sky
line at $\lambda=6258.05$ \AA. The same contamination might also
effect the \ion{Zn}{2} $\lambda2062$ line (the EW is inconsistently
lower than expected from \ion{Zn}{2} $\lambda2026$), therefore the Zn
column density is determined from \ion{Zn}{2} $\lambda2026$ only:
$\log N($\ion{Zn}{2}$) = 13.82\pm0.05$ for $b=115$ \kms.  The lack of
\ion{Fe}{2} $\lambda2586$ detection is puzzling. If, for instance, we
assume $W_r($\ion{Fe}{2}$\lambda2586) < 1$ \AA~(way above the
detection limit in the spectrum), the line is almost unsaturated, and
$\log N($\ion{Fe}{2}$)< 14.6$ for $b>70$ \kms, inconsistent with what
is found from the other \ion{Fe}{2} lines.

Fynbo et al.~(2001) report a tentative \ion{H}{1} column density
measurement of $N($\ion{H}{1}$) \approx 2\times10^{21}$ \cm.  This
leads to a relatively high metallicity with [Zn/H] $\simeq -0.13$.

\subsection{GRB010222}

The optical transient of GRB010222 at $z=1.475$ was observed with the
FLWO 1.5m telescope (Jha et al.~2001; FWHM $=6$ \AA~or $300-450$ \kms,
S/N $\simeq 10$), with the 3.58m TNG (Masetti et al.~2001; FWHM $=4.8$
\AA~ or $200-400$ \kms, S/N $=10-20$), with the 4.2m WHT (Salamanca et
al.~2001; FWHM $=3.3-5.8$ \AA~or $\sim 300$ \kms, S/N $\simeq10$), and
with Keck/LRIS and ESI (Mirabal et al.~2002; FWHM $=11-13$ \AA~or
$\sim 650$ \kms, and FWHM $=0.4-0.8$ \AA~or $\sim 30$ \kms).  Even
though the different spectra span a time interval of $\sim 27$ hours,
starting 5 hours after the burst, the absorption line equivalent
widths do not show significant time variability (Mirabal et
al.~2002). Therefore, to obtain a better estimate of column densities,
we combined the EWs of the same lines from different observations
weighted according to the errors. We constrain the effective Doppler
parameter to $b=70$ \kms, using the \ion{Si}{2} $\lambda1526$ and
\ion{Si}{2} $\lambda1808$ lines, and seven \ion{Fe}{2} lines (Figure
\ref{f1c}). We get $\log N($\ion{Si}{2}$)=16.09\pm0.05$ and $\log
N($\ion{Fe}{2}$)=15.32^{0.15}_{0.10}$. The \ion{Fe}{2} lines are
scattered in the COG diagram, probably because of non--uniform EW
measurements by the four groups.

The \ion{Zn}{2} $\lambda2026$ line at $\lambda\approx5018$ \AA~is
blended with the \ion{Mg}{1} $\lambda2026$ line.  The contamination of
\ion{Mg}{1} to this doublet is estimated using the \ion{Mg}{1}
$\lambda2852$ line at $\lambda \approx 7065$ \AA~and
$W_r($\ion{Mg}{1} $\lambda 2852) = 1.22\pm0.04$ \AA. Assuming an
effective Doppler parameter $b>35$ \kms, we get $\log N($\ion{Mg}{1}$)
< 14$ and $W_r($\ion{Mg}{1} $\lambda2026) < 0.04$ \AA. We then assume
that the feature at $\lambda\approx 5018$ \AA~ is dominated by the
\ion{Zn}{2} $\lambda2026$ absorption, with $W_r= 0.79^{+0.05}_{-0.07}$
\AA~and the lower error determined by a possible contamination from
\ion{Mg}{1} $\lambda2026$.

The \ion{Cr}{2} $\lambda2056$ EW is very low, therefore it must be
weakly or not at all saturated. From \ion{Cr}{2} $\lambda2056$ and
\ion{Cr}{2} $\lambda2066$ lines, we get $\log N($\ion{Cr}{2}$)
=14.04^{+0.04}_{-0.06}$. From this, we calculate a contribution of
\ion{Cr}{2} to the \ion{Zn}{2} $\lambda2062$ + \ion{Cr}{2}
$\lambda2062$ doublet of $W_r($\ion{Cr}{2} $\lambda2062) =0.32\pm0.04$
\AA.  For this doublet we only use the measurements given by Jha et
al., Salamanca et al.~and Mirabal et al. (weighted mean
$W_r=0.73\pm0.02$ \AA), because the spectrum by Masetti et al.~shows
strong blending with other absorption lines.  Once the \ion{Cr}{2}
contribution is taken into account, we derived $W_r($\ion{Zn}{2}
$\lambda2062)=0.41\pm0.05$ \AA. Using also the \ion{Zn}{2}
$\lambda2026$ detection, we obtain $\log N($\ion{Zn}{2}$) =
13.78\pm0.07$ for $b=70$ \kms.  The \ion{Mn}{2} column density is
obtained from the \ion{Mn}{2} $\lambda\lambda2576,2594,2606$
triplet. For $b=70$ \kms, we get $\log N($\ion{Mn}{2}$) =
13.61^{0.08}_{0.06}$.

\section{HEAVY ELEMENT ABUNDANCES IN GRB--DLAs}

\begin{figure}
\centerline{{\epsfxsize=16.cm \epsfysize=16.cm \epsfbox{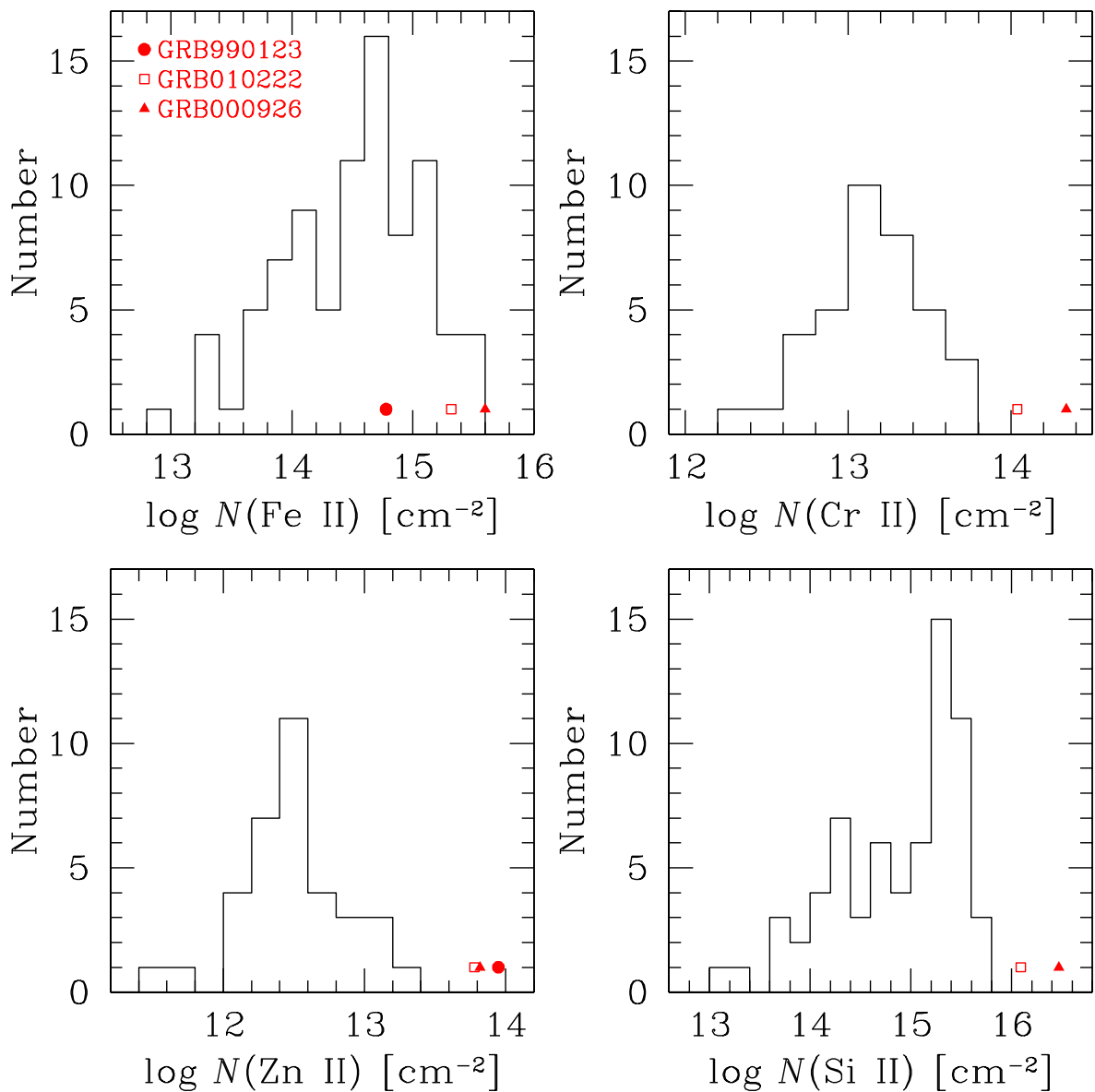}}}
\caption[f2]{Histogram column densities of \ion{Fe}{2}, \ion{Cr}{2},
\ion{Zn}{2}, and \ion{Si}{2} in QSO--DLAs. The GRB--DLA
column densities are reported as filled triangles (GRB990123),
empty squares (GRB010222) and filled circles (GRB000926).}
\label{f2}
\end{figure}

Figure \ref{f2} displays the QSO--DLA column density
histograms\footnote{We have not used \ion{Al}{2} measurements
detected in GRB000926 and GRB010222, because not included in the
QSO--DLA sample, and because [Fe/Al] $=-0.3\pm0.3$ and $-0.4\pm0.2$
for GRB000926 and GRB010222, respectively, which is not particularly
interesting for our discussion.} of \ion{Fe}{2}, \ion{Cr}{2},
\ion{Zn}{2}, and \ion{Si}{2}. These column densities are obtained
from a large compilation of QSO--DLA measurements collected from the
literature (see Savaglio 2000 for a description of the
compilation). The sample contains 98 QSO absorption line systems
associated with mostly neutral gas clouds (\ion{H}{1} column density
larger than $\sim 10^{19}$ \cm) of the ISM in $0.0 < z < 4.6$
galaxies, for which the column density of one or more of the following
elements is measured: \ion{Zn}{2}, \ion{Si}{2}, \ion{Cr}{2},
\ion{Fe}{2}.  The symbols in Figure \ref{f2} indicates the column
densities of the same ions measured for the three GRB--DLAs.  This
reveals several differences between QSO--DLAs and GRB--DLAs.  The
\ion{Zn}{2} column density in GRB--DLAs shows the most striking
difference, being much larger than in QSO--DLAs. For Si and Cr the
deviations are also very large. \ion{Fe}{2} behaves differently, being
quite consistent with the upper part of the \ion{Fe}{2} QSO--DLA
distribution.

\begin{figure}
\centerline{{\epsfxsize=17.cm \epsfysize=17.cm \epsfbox{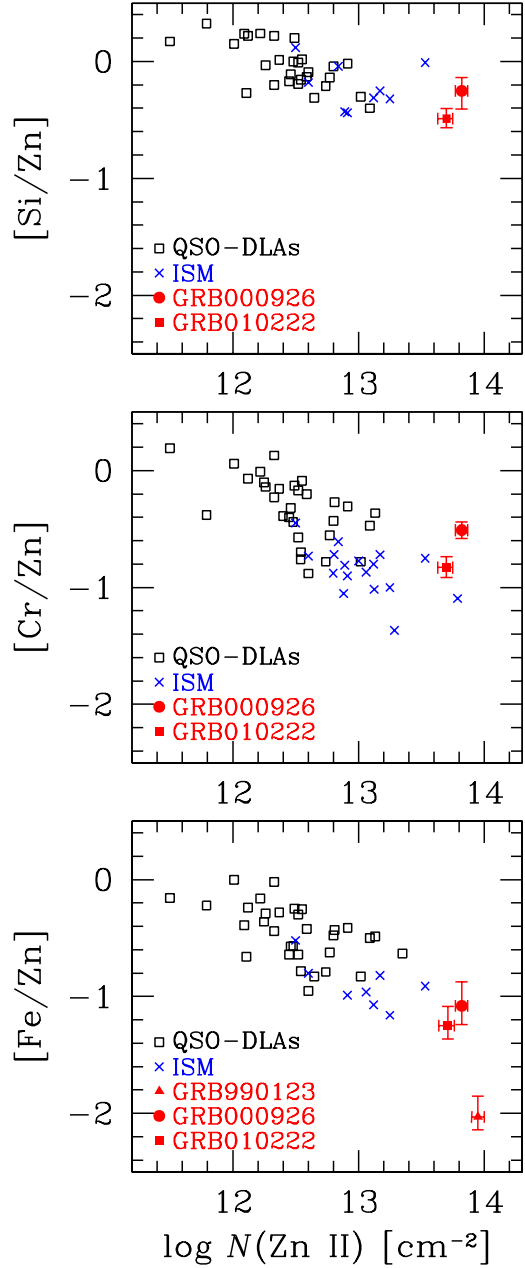}}}
\caption[f3]{Iron--to--zinc, chromium--to--zinc, and silicon--to--zinc
relative abundances, vs. \ion{Zn}{2} column density, in QSO--DLAs
(empty squares), in the ISM of the Milky Way and the Magellanic Clouds
(crosses), and in GRB--DLAs (filled symbols). Data for the local
ISM are taken from Cardelli, Sembach, \& Savage (1995), Roth \& Blades
(1995, 1997), Savage \& Sembach (1996b), Sembach \& Savage (1996),
Spitzer \& Fitzpatrick (1995), and Welty et al.~(1997, 1999).  Mean
errors for [X/Zn] in QSO--DLAs is $\sim 0.08$ dex. Errors for local
ISM column densities, when available, are typically $<0.1$ dex.  }
\label{f3}
\end{figure}

If the gas is nearly neutral, as suggested by the large EWs of the low
ionization lines, the ionization correction can be neglected and it is
possible to calculate the element relative abundance from ions with
ionization potential above 13.6 eV. This assumption is also widely
used to derive abundances in QSO--DLAs. Table \ref{t3} shows the
comparison between heavy element relative abundances in QSO--DLAs
(mean values, column four) and in the three GRB--DLAs (columns five,
six and seven) . Columns two and three also give the number of
QSO--DLAs considered and the mean redshift, respectively. For both
GRB--DLAs and QSO--DLAs, the standard solar abundances from Grevesse,
Noels, \& Sauval~(1996) are adopted. The very low [Fe/Zn] values in
three GRB--DLAs ($-2.03$, $-1.08$, and $-1.32$) are much smaller than
in QSO--DLAs: $\langle$[Fe/Zn]$\rangle = -0.46\pm0.24$ in 32
objects. This is an indication of strong dust content in the GRB
circumburst medium and/or in the ISM of the host galaxy, because while
Fe is heavily depleted on dust in the ISM of galaxies, Zn is
marginally depleted only in the densest clouds (Savage \& Sembach
1996a).  The plot of [X/Zn] abundances as a function of the
\ion{Zn}{2} column density is also very interesting (Figure
\ref{f3}). In QSO--DLAs these show a trend, the [X/Zn] being smaller
for larger $\log N($\ion{Zn}{2}).  The [X/Zn] values for GRB--DLAs are
located at the extension of the [X/Zn] vs. $\log N($\ion{Zn}{2})
distribution in QSO--DLAs.  The gap between QSO--DLAs and GRB--DLAs is
filled by a completely different class of absorbers: those in the ISM
of the MW and MC. Low values of [Fe/Zn] are also an indication of high
dust content.  Petitjean, Srianand, \& Ledoux ~(2002) discuss the gas
density in QSO--DLAs with H$_2$ detection, and suggest that the dust
depletion might be larger in denser clouds. If this can be extended to
GRB--DLAs, it might imply that their gas density is higher than in
QSO--DLAs.

The small [X/Zn] values found in GRB--DLAs might be a general property
of high metal column density systems rather than a peculiarity of GRBs
themselves. GRBs are initially brighter than QSOs, hence detectable
even in cases of heavy dust obscuration. In other words, QSO--DLAs and
GRB--DLAs might probe the same population of galaxies, but different
regions: the sight lines of the former cross moderately dense regions,
while the sight lines of the latter might cross denser regions such as
molecular clouds in a star forming environment.

As already discussed in \S~2, the COG analysis may underestimate
column densities if the lines are strongly saturated, as in the case
of \ion{Fe}{2}. This would artificially create a large dust depletion
if [Fe/Zn] were used as an indicator.  However, the underestimate of
saturated lines is in most cases not larger than 0.1 dex (Jenkins
1986). Moreover, an independent check of consistency is provided by
the comparison of \ion{Fe}{2} with \ion{Si}{2} and \ion{Cr}{2}.
\ion{Si}{2} and \ion{Fe}{2} have similar problems with saturation, but
Si is much less dust depleted (Savage \& Sembach 1996a). The
iron--to--silicon relative abundance is low in the two GRB--DLAs
discussed here, indicating high depletion of iron in dust: [Fe/Si]
$=-0.83^{+0.25}_{-0.18}$ and $-0.73^{+0.16}_{-0.11}$ in GRB000926 and
GRB010222, respectively. Also [Cr/Zn] is a dust indicator. \ion{Cr}{2}
absorption lines are generally weakly saturated, but the column
densities in the same two GRB--DLAs are low compared to \ion{Zn}{2}:
[Cr/Zn] = $-0.51 \pm0.07$ and $-0.77^{+0.08}_{-0.09}$ in the same two
GRBs.

There are indications that the \ion{H}{1} column density in GRB--DLAs
is larger than in typical QSO--DLAs. In 89 QSO--DLAs with $\log
N($\ion{H}{1}$)\geq 20.0$, only 15 (17\%) have $\log N($\ion{H}{1}$)>
21.0$.  The \ion{H}{1} column density has been estimated in GRB000301C
by Jensen et al.~(2001), who report $N($\ion{H}{1}$)
=1.5^{+3.0}_{-1.0}\times10^{21}$ \cm.  GRB000926 has also been
observed with the Nordic Optical Telescope (Fynbo et al.~2002) and a
very strong \lya~absorption line identified, with column density $\sim
2\times 10^{21}$ \cm.  This is in a very noisy part of the spectrum,
but if we assume that this \ion{H}{1} column density is close to the
correct value and neglect ionization correction, we derive [Si/H]
$=-0.38$, [Zn/H] $=-0.13$, [Cr/H] $=-0.64$, [Fe/H] $=-1.21$,
suggesting a large metallicity and dust depletion for a large
\ion{H}{1} GRB--DLA. In comparison, in 8 QSO--DLAs with $\log
N($\ion{H}{1}$)>21.0$, and measured \ion{Zn}{2} and \ion{Cr}{2}, we
find $\langle$[Zn/H]$\rangle = -1.46\pm0.34$ and
$\langle$[Cr/H]$\rangle=-1.74\pm0.22$.

\section{DUST DEPLETION AND EXTINCTION}

\begin{figure}
\centerline{{\epsfxsize=18.cm \epsfysize=18cm \epsfbox{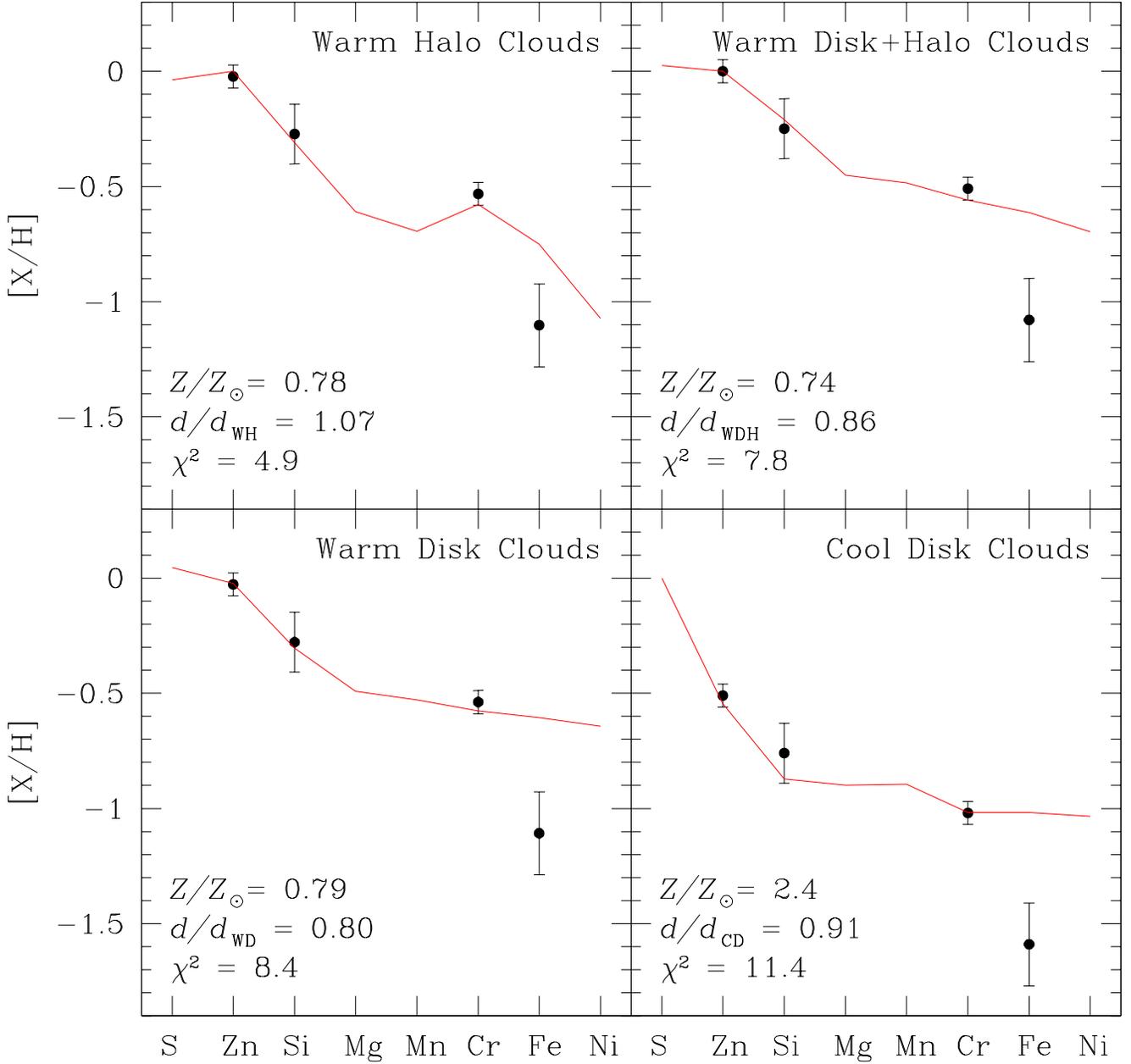}}}
\caption[f4]{Depletion patterns in the absorbing gas of GRB000926. The
models (lines) are taken from average gas--phase abundance
measurements in warm halo, warm disk+halo, warm disk, and cool disk
clouds of the Milky Way (Savage \& Sembach 1996a). The
metallicity, dust--to--metals ratio, and best--fit $\chi^2$ 
are also given.}
\label{f4}
\end{figure}

Since four elements are measured in GRB000926, it is possible to
derive its dust depletion pattern. We adopt the method described in
Savaglio (2000). Basically, we consider the four depletion patterns
observed in the Milky Way, e.g. the depletion patterns in the warm
halo (WH), warm disk+halo (WDH), warm disk (WD), and cool disk (WC)
clouds (Savage \& Sembach 1996a).  For each depletion pattern, we find
the best--fit values for the dust--to--metals ratio ($d/d_j$, where
$j$ depends on the depletion pattern assumed) and the metallicity
compared to solar $Z/Z_\sun$. The result for GRB000926 at $z=2.038$ is
shown in Figure \ref{f4}.  The best fit is given by the warm halo
cloud pattern, with $Z/Z_\sun=0.78$. This is higher than the
metallicities in QSO--DLAs: $\langle Z/Z_\sun \rangle = 0.22\pm0.23$
in 21 DLAs at $1.7 < z < 2.3$ (Savaglio 2000).

We can now estimate the obscuration of the GRB due to dust distributed
along its line of sight.  This is proportional to the column density
of metals. The extinction in the V--band, due to dust, can be
approximated by:

\begin{equation}\label{eq2}
A_V = 0.54\frac{d}{d_{WH}} \frac{Z}{Z_\sun}
\frac{N({\rm HI})}{10^{21}~{\rm cm}^{-2}}
\end{equation}

\noindent
where 0.54 is the typical value of $A_V$ in the solar neighborhood for
a column density of gas with $\log N($\ion{H}{1}$)=21.0$. For
GRB000926, we find $A_V\approx0.9$ mag.  This is much higher than in
QSO--DLAs, where $V$ extinction is typically lower than 0.1 mag.

\begin{figure}
\centerline{{\epsfxsize=18.cm \epsfysize=10.2cm \epsfbox{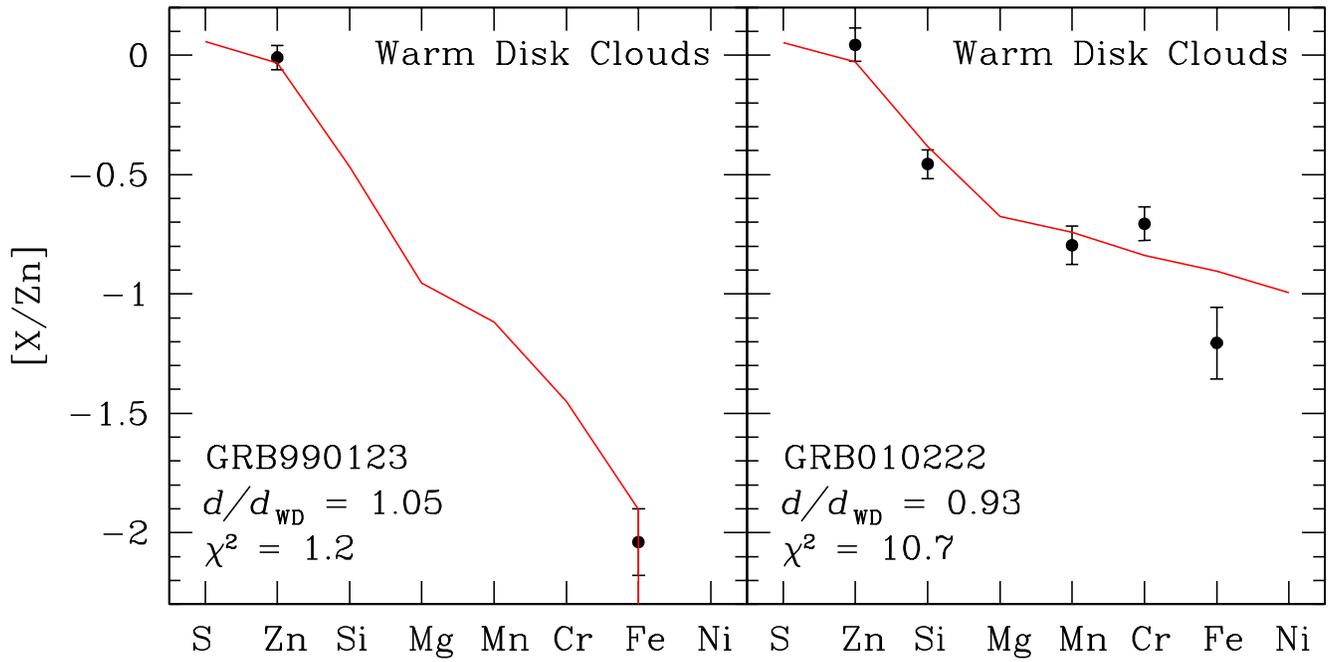}}}
\caption[f5]{Best--fit depletion patterns in the absorbing gas of
GRB990123 and GRB010222. The models (lines) are taken from average gas--phase
abundance measurements in warm disk clouds of the Milky Way (Savage
\& Sembach 1996a). Dust--to--metals ratio and best--fit $\chi^2$ 
are also given.}
\label{f5}
\end{figure}

The depletion pattern and extinction can also be calculated for
GRB990123 and GRB010222, because this depends on the product of
metallicity and \ion{H}{1} column density, i.e., the column density of
metals. We consider the \ion{Zn}{2} column density and neglect the
dust depletion correction for this element.  Best--fits for these two
GRB--DLAs give a warm halo cloud depletion pattern (Figure \ref{f5}),
and dust extinctions of $A_V\approx1.1$ and 0.7 mags for GRB990123 and
GRB010222, respectively.

\section{DISCUSSION}

We have used the curve of growth to study the heavy element absorption
lines associated with three GRBs, namely GRB990123 (Kulkarni et
al.~1999), GRB000926 (Castro et al.~2001), and GRB010222 (Masetti et
al.~2001; Jha et al.~2001; Salamanca et al.~2001; Mirabal et
al.~2002). These are the only GRBs with available UV spectra obtained
so far with quality good enough to detect high S/N absorption
lines. All GRB spectra show very high \ion{Zn}{2} absorption
associated with the circumburst medium or interstellar medium of the
host galaxy, from which we derive a mean column density of
$\langle\log N($\ion{Zn}{2}$)\rangle=13.85$ at $\langle z \rangle
=1.70$. This is $\sim3$ times larger than the largest \ion{Zn}{2}
column density detected so far in QSO--DLAs.  High \ion{Zn}{2} columns
indicate high metallicity and/or low ionization of the gas, also
supported by the presence of strong \ion{Fe}{2}, \ion{Si}{2}, and
\ion{Cr}{2}, as well as \ion{Mg}{2} together with \ion{Mg}{1}.
Moreover, the GRB--DLAs analyzed here satisfied the criteria found by
Rao \& Turnshek (2000) that are used for QSO--DLA selection, namely
$W_r($\ion{Mg}{2} $\lambda2796) \geq 0.6$ \AA~and $W_r($\ion{Fe}{2}
$\lambda2600) \geq 0.5$ \AA.

Since the metal column densities are so high, the dust extinction can
be very important. A good indicator of dust content is given by the
[Fe/Zn] values; in the GRB--DLAs these are much smaller (higher dust
depletion) than those seen in QSO--DLAs:
$\langle$[Fe/Zn]$\rangle=-1.48$ in the former,
$\langle$[Fe/Zn]$\rangle=-0.46$ in the latter. The combination of
large column densities of metals and large dust depletion can lead to
large extinction in GRB--DLAs. If we compare metal column densities
with the ones in the solar neighborhood, where extinction is very well
studied, we find optical extinction of $A_V \approx 0.7, 0.9$ and 1.1
magnitudes. This is much larger than that found for most QSO--DLAs:
$A_V\lesssim0.1$.  We notice, however, that another possible
explanation for the very low [Fe/Zn] values would be a nucleosynthetic
production of Fe and Zn in GRBs and/or their environment being different
from the standard solar pattern.

These strong differences between QSO--DLAs and GRB--DLAs do not
necessarily indicate a difference in the nature of the host galaxies.
Indeed, the distribution of the Zn relative abundances as a function
of \ion{Zn}{2} column density in GRB--DLAs and QSO--DLAs is consistent
for these two classes of objects. [Fe/Zn], [Cr/Zn], and [Si/Zn]
decrease as a function of $\log N($\ion{Zn}{2}) for QSO--DLAs and
GRB--DLAs, which are located in the upper left part and in the lower
right part of the diagram, respectively.  The QSO--DLA and
GRB--DLA distributions are bridged by the heavy element abundances in
the diffuse and dense ISM of the Milky Way and the Magellanic
Clouds. This suggests that the galaxies associated with the absorbing
gas around GRBs are not different from those of QSO--DLAs.  The
important difference is in the column densities of the intervening
ISM: GRBs are (at least for some time) much brighter than optically
selected QSOs, so their sight lines can probe much denser and dustier
regions of galaxies. Indications of a bias effect in QSO--DLAs due to
dust obscuration have been pointed out by several authors (Fall \& Pei
1993; Boiss\'e et al.~1998; Savaglio 2000; Petitjean et al.~2002). 

Reichart (1998), Jensen et al.~(2001), Rhoads \& Fruchter (2001),
Fynbo et al.~(2001), and Galama \& Wijers (2001) have estimated the
dust extinction in GRBs from curvature in their spectra, assuming that
the intrinsic GRB spectra from the UV to IR bands are power laws.
This approach gives a small $V$ extinction only if MW or MC laws are
assumed: $A_V\approx 0-0.2$ for GRB970508, GRB000301C, GRB000926, and
GRB990123. The result for GRB990123 is supported by the lack of
spectral slope variation going from the optical to x--ray bands
(Castro-Tirado et al.~1999).  \v{S}imon et al.~(2001) have found that
the color variation during GRB decline in 17 objects (three of which
are the ones studied in this paper) is relatively small, suggesting
similar and low reddening in these GRBs. However, they also notice
that another sample of five GRBs (not included in the first one, for
lack of complete photometric information) have much redder colors.
Low extinction can be explained by sublimation of dust grains due to
UV emission (Waxman \& Draine 2000). Fruchter, Krolik, \& Rhoads
(2001) suggest grain destruction by GRB heating and grain charging.
These effects can be present up to a distance of $\sim100$ pc from the
GRB source.

This is in principle in contradiction with our findings. However, the
dust disruption mechanisms suggested by Fruchter et al.~(2001), or
other mechanisms, such as sputtering (Draine \& Salpeter 1979),
predict that small grains are destroyed first. Dust destruction of
small grains would still give a large content of dust, if the initial
grain--size distribution is dominated by large grains, as suggested in
models of MW an MC extinction curves (Weingartner \& Draine
2001). Therefore, what remains in dust can still be an important
fraction of the total metal content.  Grain coagulation favored by
high densities, leading to the formation of large grains, is also
possible (Mathis 1990). In these cases, MW or MC extinction curves
might be inappropriate to describe dust obscuration. A dust grain
distribution skewed toward large grains would give a grey extinction,
therefore may explain the low reddening found in GRB afterglows
(Stratta et al.~2002). This is a plausible speculation, but
quantitative calculations, needed to determine if it works in
practice, are beyond the scope of this paper. A first attempt is
instead presented by Perna \& Lazzati (2002), who have developed a
detailed numerical code that describes, among other effects, the dust
destruction under an intense X--ray and UV radiation field the very
first seconds after the burst.  Alternative explanations are unknown
systematic effects. For instance, the low reddening suggested relies
on the assumption of an intrinsic power law emission of the optical
afterglow, which in principle cannot be tested.  It might also be that
the extinction in itself is a power law, and when combined with the
intrinsic GRB power law emission, would still give a power law
spectrum. Therefore, this would require low or no rectification.

The main result of the present work is the strong difference between
QSO--DLA and GRB--DLA heavy element column densities. GRB--DLAs occur
most likely along the sight line of much brighter sources, and are
embedded in star forming and therefore metal rich regions, where the
dust extinction is probably not negligible.  This is another evidence
supporting the idea that DLAs in front of the fainter QSO radiation,
can be detected and studied in detail only when the sight line crosses
less dense, metal polluted and/or extincted regions of high redshift
galaxies.

\acknowledgments We thank Tim Heckman, Fiona Harrison, Eliana Palazzi
and Ken Sembach for fruitful information on GRB spectra and ISM
absorptions. We are particularly grateful to Daniela Calzetti, Julian
Krolik, Nicola Masetti, Nino Panagia, and James Rhoads for stimulating
discussions that helped to improve this paper.

\clearpage

\clearpage

\clearpage

\clearpage

\clearpage

\clearpage

\clearpage

\clearpage

\begin{table}
\caption[t1]{Rest--frame equivalent widths}\label{t1}
\begin{center} 
\begin{tabular}{lcccc} 
\tableline\tableline&&&\\[-10pt] 
&& \multicolumn{3}{c}{$W_r$ (\AA) } \\
\cline{3-5}\\[-7pt] 
     & $f_\lambda$ & GRB990123 & GRB000926 & GRB010222 \\
Line &            & $(z=1.6004)$ & $(z=2.038)$ & $(z=1.475)$ \\ 
[5pt]\tableline&&&&\\[-5pt] 
\ion{Zn}{2} $\lambda2026$ & 0.489 & $0.800\pm0.065$ & $0.900\pm0.067$ & $0.79^{+0.05}_{-0.07}$ \tablenotemark{c,d} \\
\ion{Zn}{2} $\lambda2062$ & 0.256 & $0.485\pm0.038$\tablenotemark{a} & $<0.28$\tablenotemark{b} & $0.41\pm0.05$\tablenotemark{c,e} \\
\ion{Si}{2} $\lambda1526$ & 0.127 &   -- & $2.64\pm0.18$ & $1.53\pm0.09$\tablenotemark{f} \\
\ion{Si}{2} $\lambda1808$ & 0.00218 & -- & $1.130\pm0.089$ & $0.55\pm0.02$\tablenotemark{f} \\
\ion{Cr}{2} $\lambda2056$ & 0.105 & -- & $0.690\pm0.050$ &  $0.35\pm0.03$\tablenotemark{g} \\
\ion{Cr}{2} $\lambda2066$ & 0.0515 & -- & $<0.2$ &  $0.31\pm0.08$\tablenotemark{g} \\ 
\ion{Cr}{2} $\lambda2062$ & 0.078 & -- & $<0.28$\tablenotemark{b} &  $0.32\pm0.04$\tablenotemark{c,e}   \\
\ion{Fe}{2} $\lambda1608$ & 0.058 & -- & $2.15\pm0.16$ & $0.47\pm0.04$\tablenotemark{g} \\
\ion{Fe}{2} $\lambda2260$ & 0.00244 & -- & -- & $0.52\pm0.18$\tablenotemark{h} \\
\ion{Fe}{2} $\lambda2344$ & 0.114 & $1.17\pm0.11$ & $3.41\pm0.24$ & $1.92\pm0.04$\tablenotemark{i} \\
\ion{Fe}{2} $\lambda2374$ & 0.0313 & -- & $2.57\pm0.20$ & $1.60\pm0.06$\tablenotemark{j}  \\
\ion{Fe}{2} $\lambda2382$ & 0.320 & $1.38\pm0.15$ & $3.07\pm0.31$ & $2.35\pm0.06$\tablenotemark{j}\\
\ion{Fe}{2} $\lambda2586$ & 0.0691 & $1.09\pm0.14$ & -- & $1.53\pm0.05$\tablenotemark{f} \\
\ion{Fe}{2} $\lambda2600$ & 0.239 & $1.338\pm0.054$ & $3.65\pm0.26$ & $2.31\pm0.06$\tablenotemark{f} \\ 
\ion{Mn}{2} $\lambda2576$ & 0.3508 & -- & -- &  $0.60\pm0.05$\tablenotemark{f} \\
\ion{Mn}{2} $\lambda2594$ & 0.2710 & -- & -- & $0.68\pm0.09$\tablenotemark{c} \\
\ion{Mn}{2} $\lambda2606$ & 0.1927 & -- & -- & $0.56\pm0.12$\tablenotemark{h} \\
[2pt]\tableline
\end{tabular}
\tablenotetext{a}{\ion{Cr}{2} $\lambda2062$ contribution to the \ion{Zn}{2} $\lambda2062$ + \ion{Cr}{2}$\lambda2062$ blend is negligible.}
\tablenotetext{b}{Noisy region of the spectrum, given only upper limit based on the total EW + $3\sigma$, reported by Castro et al.~2001.} 
\tablenotetext{c}{Weighted mean from Jha et al.~2001, Salamanca et al.~2001, and Mirabal et al.~2002.}
\tablenotetext{d}{Lower error found assuming possible contamination from \ion{Mg}{1} $\lambda2026$ line.} 
\tablenotetext{e}{Contribution of the \ion{Cr}{2} $\lambda2062$ line to the 
\ion{Zn}{2} $\lambda2062$ + \ion{Cr}{2}$\lambda2062$ blend calculated using the \ion{Cr}{2} column density as derived from \ion{Cr}{2} $\lambda\lambda2056,2066$.}
\tablenotetext{f}{Weighted mean from Masetti et al.~2001, Salamanca et al.~2001, Jha et al.~2001, and Mirabal et al.~2002.} 
\tablenotetext{g}{Weighted mean from Salamanca et al.~2001 and Mirabal et al.~2002.}
\tablenotetext{h}{Mirabal et al.~2002.}
\tablenotetext{i}{Weighted mean from Masetti et al.~2001, Jha et al.~2001, and Mirabal et al.~2002.} 
\tablenotetext{j}{Weighted mean from Jha et al.~2001 and Mirabal et al.~2002.}
\end{center}
\end{table}

\clearpage

\begin{table}
\caption[t2]{Column densities}\label{t2} 
\begin{center} 
\begin{tabular}{lccc}
\tableline\tableline&&&\\[-10pt]
& \multicolumn{3}{c}{$\log N(X)$ (\cm)} \\
\cline{2-4}\\
[-7pt] 
Ion & GRB990123 & GRB000926 & GRB010222 \\
\tableline&&&\\[-8pt]
\ion{Zn}{2} & $13.95^{+0.05}_{-0.05}$ & $13.82^{+0.05}_{-0.05}$ & $13.78\pm0.07$ \\
\ion{Si}{2} & -- & $16.47^{+0.10}_{-0.15}$ & $16.09\pm0.05$ \\
\ion{Cr}{2} & -- & $14.34^{+0.05}_{-0.05}$ & $14.04^{+0.04}_{-0.06}$ \\
\ion{Fe}{2} & $14.78^{+0.17}_{-0.10}$ & $15.60^{+0.20}_{-0.15}$ & $15.32^{+0.15}_{-0.10}$ \\
\ion{Mn}{2} & -- & -- & $13.61^{+0.08}_{-0.06}$ \\
[2pt]\tableline
\end{tabular}
\end{center}
\end{table}

\clearpage

\begin{table}
\caption[t3]
{Element relative abundances in QSO--DLAs and GRB--DLAs}{\label{t3}}
\begin{center}
\begin{tabular}{lccccccc}
\tableline\tableline&&&&&&&\\[-10pt] 
& \multicolumn{3}{c}{QSO--DLAs} & & \multicolumn{3}{c}{GRB--DLAs} \\
\cline{2-4}\cline{6-8} \\
[-10pt]
$[$X/Zn$]$  & No.  & $\langle z \rangle$ & $\langle$[X/Zn]$\rangle$ & & GRB990123 & GRB000926 & GRB010222 \\
[2pt]\tableline&&&&&&&\\[-8pt] 
$[$Fe/Zn$]$ & 32 & 1.994 & $-0.46\pm0.24$ & & $-2.03^{+0.18}_{-0.11}$ & $-1.08^{+0.21}_{-0.16}$ & $-1.32^{+0.17}_{-0.12}$ \\
$[$Si/Zn$]$ & 28 & 2.064 & $-0.04\pm0.19$ & & -- & $-0.25^{+0.11}_{-0.16}$ & $-0.59^{+0.09}_{-0.09}$ \\
$[$Cr/Zn$]$ & 30 & 2.013 & $-0.32\pm0.28$ & & --  & $-0.51^{+0.07}_{-0.07}$ & $-0.77^{+0.08}_{-0.09}$ \\
$[$Mn/Zn$]$ & 11 & 1.392 & $-0.59\pm0.26$ & & -- & -- & $-0.91^{+0.11}_{-0.09}$ \\
[2pt]\tableline
\end{tabular}
\end{center}
\end{table}

\end{document}